\documentclass[sigconf,language=english]{acmart}
\AtBeginDocument{%
  }

\setcopyright{acmcopyright}
\copyrightyear{2024}
\acmYear{2024}
\acmDOI{XXXXXXX.XXXXXXX}
\acmConference[ICSE 2024]{46th International Conference on Software Engineering}{April 2024}{Lisbon, Portugal}
\acmPrice{15.00}
\acmISBN{978-1-4503-XXXX-X/18/06}

\begin{document}

\title{Detecting Android Malware: From Neural Embeddings to Hands-On Validation with BERTroid}

\author{Meryam Chaieb}
\email{meryam.chaieb.1@ulaval.ca}
\affiliation{%
  \institution{Laval university}
  \city{Quebec}
  \country{Canada}
}

\author{Mostafa Anouar Ghorab}
\email{mostafa-anouar.ghorab.1@ulaval.ca}
\affiliation{%
  \institution{Laval university}
  \city{Quebec}
  \country{Canada}
}

\author{Mohamed Aymen Saied}
\email{mohamed-aymen.saied@ift.ulaval.ca}
\affiliation{%
  \institution{Laval university}
  \city{Quebec}
  \country{Canada}
}

\begin{abstract}
As cyber threats and malware attacks increasingly alarm both individuals and businesses, the urgency for proactive malware countermeasures intensifies. This has driven a rising interest in automated machine learning solutions. Transformers, a cutting-edge category of attention-based deep learning methods, have demonstrated remarkable success. In this paper, we present BERTroid, an innovative malware detection model built on the BERT architecture. Overall, BERTroid emerged as a promising solution for combating Android malware. Its ability to outperform state-of-the-art solutions demonstrates its potential as a proactive defense mechanism against malicious software attacks. Additionally, we evaluate BERTroid on multiple datasets to assess its performance across diverse scenarios. In the dynamic landscape of cybersecurity, our approach has demonstrated promising resilience against the rapid evolution of malware on Android systems. While the machine learning model captures broad patterns, we emphasize the role of manual validation for deeper comprehension and insight into these behaviors. This human intervention is critical for discerning intricate and context-specific behaviors, thereby validating and reinforcing the model's findings. 
\newline Additional technical details are available in
the replication packages \footnote{}.

\end{abstract}



\maketitle

\section{Introduction}
Information technology security is an essential component of the digital ecosystem and it continues to attract the attention of researchers in both academia and business. The phenomenal popularity of the Android operating system, which accounts for a significant 40.17\%  part of the global Operating System (OS) market \footnote{http://surl.li/jnymt} and an astonishing 70.79\%  share of the mobile phone OS \footnote{http://surl.li/dwjdr} emphasises the need of improving this operating system security.
\newline The broad acceptance of the Android operating system and the fast proliferation of smartphones have transformed how people engage with technology. While Android smartphones are an essential part of our everyday lives, they are also a target for cybercriminals aiming to exploit vulnerabilities and distribute malware. The rise of mobile malware threatens user privacy, data security and the entire integrity of the Android ecosystem.
As a result, detecting and preventing malware Android apps has emerged as a serious problem in the realm of mobile security.

Statistics show a significant increase in the number of dangerous Android applications, with countless malware types launching on a daily basis \cite{ref20_kalpana_malware_2023}. Furthermore, third-party app stores and unauthorized sources continue to be fertile ground for the dissemination of malicious apps, increasing the danger for Android users. As of 2022, the Google Play Store alone held hundreds of fraudulent apps that currently evade Google Bouncer’s detection technology \cite{ref3_rahman_search_2017} with over 2 million downloads of apps that contained harmful code, allowing them to circumvent security protections and infiltrate unsuspecting users' devices \cite{ref4_noauthor_dangerous_nodate}. 

Malware detection is an essential component in protecting IT systems from potential threats. In response to the growing menace of Android malware, researchers and security experts have developed a wide array of detection techniques. These include static analysis-based techniques, dynamic analysis-based techniques, machine learning-based approaches, behavior-based detection and hybrid methods. Static analysis scrutinizes the code structure of Android applications, while dynamic analysis observes the app's runtime behavior to detect suspicious activities. Machine learning models leverage vast datasets to identify malware patterns, while behavior-based detection monitors app behavior for anomalous actions. Combinations of these methods offer promising solutions, but each has its strengths and limitations.


In this study, we introduce a transformer-based machine learning approach, the results of which are validated using a manual protocol. This strategy addresses the continually evolving threats related to permissions, essential for safeguarding user privacy expectations \cite{red26_7965360}. BERT excels beyond traditional AI techniques in various facets of malware detection. Distinct from conventional methods, our approach uses a manual validation process that incorporates both static and dynamic analyses to robustly confirm model outcomes. We assess our method's efficacy on renowned datasets like MalDozer, AndroZoo, and Drebin. Beyond just pinpointing current malicious activities, we evaluate how our technique adapts to changes in Android app permissions, offering a comprehensive appraisal of its real-world scenarios.
\newline Our proposed approach for detecting malware in Android applications offers several significant contributions:
\begin{itemize}
    \item Our research unveils a unique approach to Android malware detection that hinges exclusively on application permissions, enhanced by the capabilities of the BERT architecture. This methodology stands out for its outstanding efficacy and its ability to adjust to changing permissions.
    \item We present a manual protocol for malware detection. This process promotes deeper comprehension of app behaviors and serves as a reference to validate the results of our automated solution, reinforcing its reliability and consistency.
\end{itemize}

The rest of this paper is organized as follows: Essential background definitions and context are provided in Section \ref{sec2} to lay the foundation for our suggested strategy. An in-depth review of relevant research in the field of Android malware detection is provided in Section \ref{sec3}. The whole methodology of our approach and its many steps are described in Section \ref{sec4}. The main research challenges that served as the basis for our evaluation are discussed in Section \ref{sec5} along with the findings of our tests. The validity of our study is critically examined in Section \ref{sec6}. Section \ref{sec7} brings the work to a close by summarizing the major discoveries and learnings from our research and exploring prospective directions for further investigation.

\section{Background}\label{sec2}

It has been 19 years since the first mobile malware known as "Mosquito" was developed by the "Ojam" company \cite{key:article}. Six years later, with the growing popularity of the Android OS, malware developers shifted their attention to it, leading to the discovery of the first Android malware, DroidSMS. DroidSMS operates by sending SMS to premium rate services owned by the attacker, charging users higher fees than standard phone numbers. This malicious behavior requires the SEND\_SMS permission. It's important to note that any behavior in the Android system requires one or more special permissions, which serves as the basis for our approach designed to detect malware in Android applications. 

In the Android system, permissions play a crucial role in ensuring the security and confidentiality of user data and are instrumental in identifying behaviors that could jeopardize user security and privacy \cite{red26_7965360}. In the latest versions of Android, starting from version 6.0 \footnote{https://developer.android.com/training/permissions/requesting?hl=fr}, the permissions requested during app installation reveal the actions an application may take \cite{ref21_wang_runtime_2021} \cite{ ref22_wang_aper_2022}, prior to this version, apps declared permissions in their manifest file were granted permissions during installation \cite{ref23_bogdanas_dperm_2017}. Security experts meticulously analyze these permissions for any unusual or unlawful activities that could indicate the presence of malware. Excessive or inappropriate permissions, such as unauthorized access to contacts or communications, may raise red flags and warrant further investigation.

\textbf{ Natural language processing (NLP) } offers new possibilities for improving security measures. We can use the power of modern language models to analyze the textual components of Android applications, such as permissions, approaching malware detection as a natural language processing challenge. This paradigm shift offers a number of important benefits, including improved understanding of context, resistance to evasion strategies and efficient analysis of huge datasets. The multifaceted and frequently changing nature of malware has posed difficulties for traditional AI-based detection strategies, such as signature and rule-based approaches.
\textbf{Transformers} represent a major advance in natural language processing tasks, and BERT is an impressive example of this technology \cite{ref6_devlin_bert_2019}. The fundamental components of transformers are self-attention mechanisms, which enable the model to pay attention to several words in a sentence at once while learning their connections. Transformers can successfully capture long-range interactions and contextual information through their bidirectional mechanism, overcoming the drawbacks of conventional sequential models such as RNNs and LSTMs \cite{ref7_vaswani_attention_2017}.
The ability of transformers to contextualize words in respect of their surroundings is at the origin of BERT's success. The entire context of a word cannot be taken into account in a single pass, as traditional language models analyze text sequentially, unlike BERT, which improves language comprehension and the overall performance of NLP tasks \cite{ref6_devlin_bert_2019}.

\textbf{BERT (Bidirectional Encoder Representations from Transformers)}, which can capture bidirectional context and offer in-depth information on the complex interactions between textual parts, appears as a revolutionary model in this scenario. By leveraging BERT's contextual language comprehension for generating embeddings, we can significantly enhance the accuracy and efficiency of malware detection methods.
Unlike traditional strategies that rely on fixed patterns or signatures, its understanding of context enables it to capture semantic connections and contextual nuances making it easier to identify complex patterns and variations in malware behavior \cite{ref5_loureiro_analysis_2021}. This advantage makes BERT resistant to evasion tactics, ensuring its flexibility to adapt to changes in malware penetration techniques used by malicious applications. In addition, BERT's pre-trained language representations enable large datasets to be processed efficiently, enabling fast and proactive analysis of a variety of Android applications. The scalability of traditional signature and rule-based systems is limited for large-scale malware research, as they often require intensive computing resources and human modifications to the rules. Also, BERT's transfer learning capability enables accurate models to be developed with less data, boosting the efficiency and scalability of the malware detection system. Actually, transfer learning is facilitated by its ability to learn from a large corpus of textual data through pre-training. This means it can learn general language properties and then be refined for particular tasks such as malware detection \cite{ref6_devlin_bert_2019}.

\section{Related work}\label{sec3}
\paragraph{}
Over the past decades, software engineering research identified and attempted to solve a variety of issues pertaining to several phases of the software lifecycle. 
However, the fast pace of evolution in the IT industry  and the staggering growth of new technologies \cite{vayghan2021kubernetes} 
based on APIs \cite{saied2015could, shatnawi2018identifying, mujahid2021toward}, containers \cite{vayghan2019kubernetes}, 
microservices \cite{sellami2022improving, almarimi2019web, sellami2022hierarchical, saidani2019towards}, 
cloud and virtualization, put an increasing pressure on software development \cite{benomar2015detection} and deployment \cite{vayghan2019microservice, vayghan2018deploying} 
practice to fully exploit this paradigm shift. This led to constant questioning of existing techniques \cite{saied2015could1} and results of software 
engineering research \cite{saied2020towards, saied2018improving}, leading to investigating the use of AI and ML-based techniques to solve software engineering problems 
in topics related to software reuse \cite{gallais2020api}, recommendation systems \cite{saied2016automated}, mining software repositories \cite{saied2020towards}, 
software data analytics and patterns mining \cite{saied2018towards, huppe2017mining, saied2016cooperative, saied2015mining} , 
program analysis and visualization \cite{saied2015observational, saied2015visualization}, testing in the cloud environment, Edge-Enabled systems \cite{mouine2022event}, 
microservices architecture \cite{sellami2022combining} and mobile applications security. 

\paragraph{}

Faced with the growing threat posed by malicious applications attacking the Android operating system, research efforts in the field of Android malware detection have increased considerably in recent years. This section offers a look at the different techniques and approaches used by researchers to identify malware on Android by classifying them into several categories:

\textbf{Static analysis based-techniques } dive into the essential components of Android applications, including their permissions and code structures. One such approach, MalDozer \cite{ref8_karbab_maldozer_2018}, uses deep learning techniques for sequence classification. By analyzing the raw sequence of API method calls in apps, MalDozer automatically extracts and learns patterns indicative of malicious and benign behavior, enabling it to identify Android malware.
Another remarkable approach introduced by Rahali et al.\cite{ref9_9659287} 
 involves extracting information from the application manifest file, including a list of permissions, activities, services, broadcast receivers, content providers, versions, and metadata. Souani and al.\cite{w3} in their research undertake two empirical studies: (1) it reproduces MalBERT's results using BERT with Android manifests from 265k apps, and (2) explores BERT's capability in classifying malware families. By exploiting this information, the authors successfully identify potential instances of malware in Android applications. AndroVul \cite{ref13_zakeya_probing_2022} serves as a repository for Android security vulnerabilities based on permissions, code smells and security artifacts generated from Androbug.
 The research delves into the application of diverse machine learning techniques and classifiers to proficiently categorize malicious apps.
 Apposcopy \cite{ref24_10.1145/2635868.2635869} focuses on identifying classes that steal user privacy, while \cite{ref25_feng_automated_2017} utilizes semantic malware signatures. ArgusDroid \cite{ref27_bai_argusdroid_2023} employs mining of the permission-API knowledge graph to detect Android malware variants. Study \cite{ref28_15-5-2-942-5052} utilizes a clone detection technique to identify similar source files in malicious applications. Additionally, \cite{ref30_10.1145/2897073.2897707} introduces the MassDroid tool, which collects security-relevant operation traces from the call graph and applies data-flow analysis to achieve a classification using supervised learning.
In the study by Sharma et al.\cite{ref10_sharma_investigation_2018}, opcodes are utilized as the primary feature for malware detection, and various classifiers are analyzed. Sachdeva et al.\cite{ref11_sachdeva_android_nodate}, on the other hand, propose a machine learning-based approach for classifying Android applications into three classes: safe, suspicious, and highly suspicious. They extract and select 35 features from the Mobile Security Framework, which are obtained through penetration testing. Study \cite{w2} introduces a multi-view deep learning approach for Android malware detection, emphasizing zero-day threats. Instead of relying on traditional domain knowledge, it utilizes raw opcodes, app permissions, and Android API package usage. 
\newline Although static analysis has the advantage of being fast and scalable, its limitations in terms of sophisticated obfuscation methods and its inability to uncover the dynamic behavior of packaged or encrypted malware have prompted researchers to investigate other complementary techniques, such as dynamic analysis.

\textbf{Dynamic analysis based-techniques} is of great importance in the field of Android malware detection, as it enables the behavior of a running application to be observed and monitored in real time, one of these notable contributions is Android X-Ray \cite{ref14_Karthikeyan2022AndroidX}. It extracts a wealth of information about an application's behavior, enabling machine-learning models to be trained and giving security analysts the means to make informed decisions quickly. Similarly, the DynaMalDroid framework \cite{ref15_HaidrosRahimaManzil2022DynaMalDroidDA} presents a holistic approach encompassing dynamic analysis, feature engineering and detection modules, enhancing malware detection capabilities. Oak and al in their research \cite{w1} delved into Android malware detection using deep learning on dynamic analysis of app activity sequences. Despite utilizing a highly imbalanced dataset, their BERT-based model achieved an F1 score of 0.919, demonstrating significant advancement over existing methods.
In conclusion, dynamic analysis has emerged as an indispensable tool in malware detection, offering a deep dive into the real-time behaviors of applications. Yet, while it provides a robust line of defense by unveiling potential malicious activities, it isn't without its limitations. It often require more computational resources, given the need to execute the applications in real or emulated environments. This process can be time-consuming and might not guarantee the triggering of all malicious behaviors during the analysis phase. Furthermore, sophisticated malware can detect when it's being observed in such environments, thereby behaving benignly to avoid detection. The future of a secure Android ecosystem hinges on blending the strengths of both dynamic and static analysis, ensuring a holistic approach that can adapt to the evolving landscape of cyber threats.

\textbf{Hybrid analysis based-techniques}, which combine the power of both static and dynamic techniques, has emerged as a critical approach in the field of malware detection due to its importance in improving the efficiency and accuracy of identifying harmful activity. Study \cite{ref16_Bokolo2022HybridAB} proposes a cross-inspection framework that allows for concurrent static and dynamic detections, resulting in better performance. The study \cite{ref17_Hadiprakoso2020HybridBasedMA} uses hybrid analysis on separate static and dynamic datasets, extracts combined features, and achieves a 5\% increase in detection rates.

In wrapping up our review of related literature, the transition from static to dynamic and ultimately to hybrid analysis highlights the ongoing evolution in Android malware detection.

\section{Proposed approach}\label{sec4}
\begin{figure*}
    \centering {\includegraphics [width=15cm,height=44mm]{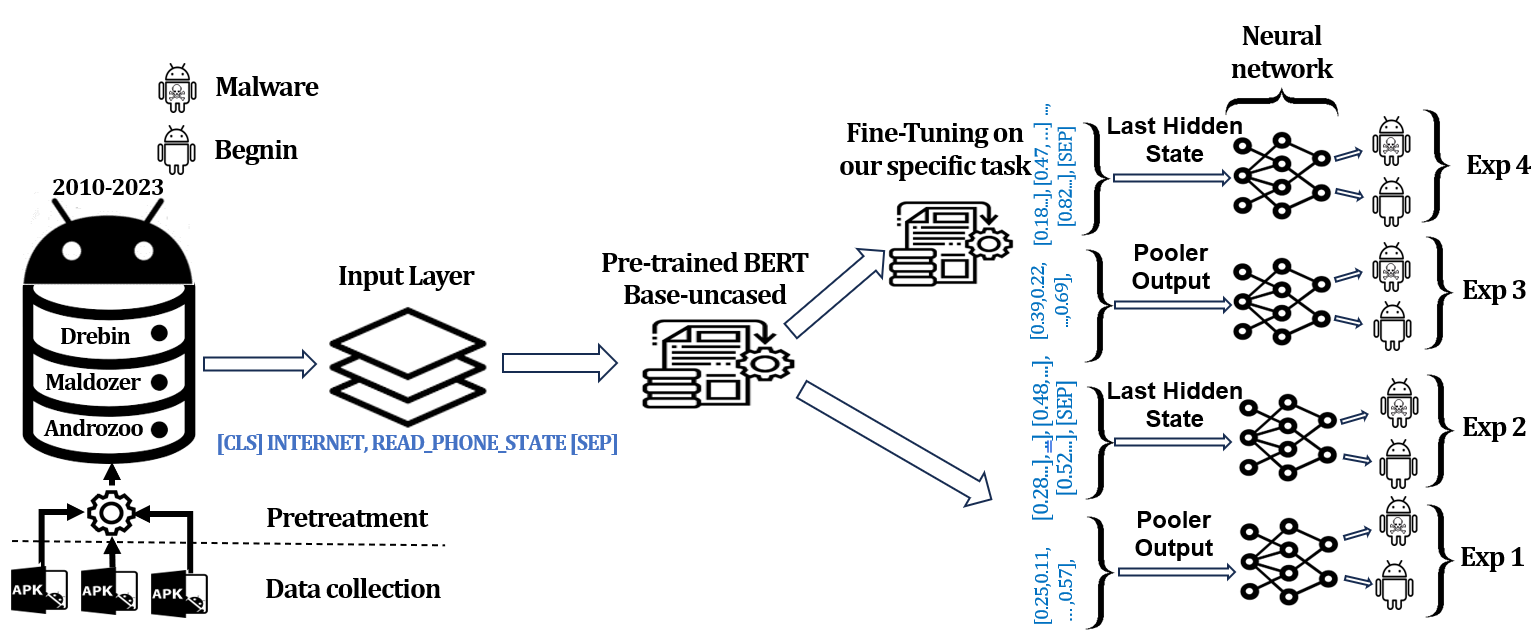}}
    \caption{\small Approach Diagram Summary }
    \label{figappr}
\end{figure*}
In response to concerning results from the Google Play Protect system \footnote{https://support.google.com/googleplay/answer/2812853?hl=} reported in the study \cite{ref31_10.1145/3510003.3510161}, which detected only 91.3\% of threats in a recent AV-Test report \footnote{https://www.av-test.org/en/antivirus/mobile-devices/android/may-2023/google-play-protect-35.6-233310/} (May 2023) compared to the industry standard of 97.3\% with a protection score of 3.5/6, our research aims to pioneer improved malware detection techniques. As depicted in Figure \ref{figappr}, we began by extensively collecting APK files to build our dataset. This dataset was processed to construct embeddings using the BERT architecture, which then trained and fine-tuned the neural network for malware app classification. Further details on the methodology and the designed experiments will be outlined in the upcoming sections.

\subsection{Data collection}
Selecting appropriate datasets is crucial for the integrity and assessment of our proposed system.
Drebin\cite{ref19_arp_drebin_2014} is a renowned repository in Android malware detection. The Drebin dataset \footnote{https://www.sec.tu-bs.de/~danarp/drebin/download.html} encompasses 5,560 malware app files and 123,453 benign applications, collected between August 2010 and October 2012. Though Drebin offers an expansive collection, it has a limited set of malware apps.
This limitation prompted us to integrate the Maldozer dataset\cite{ref8_karbab_maldozer_2018}, distinguished for its validated malware samples. Specifically, the Maldozer dataset contains 20,098 authenticated malware apps.
Drebin and Maldozer, despite their robustness, did not provide a time-based perspective, which is crucial when investigating the evolution of Android permissions over time. This gap steered us towards Androzoo\cite{ref18_Allix:2016:ACM:2901739.2903508}. The Androzoo repository \footnote{https://androzoo.uni.lu/} houses an impressive collection of over 22,025,078 apps as of the date of download. We utilized its main CSV file, which offers essential information about each application: hash keys (sha256, sha1, md5), size data, binary date, package name, version code, marketplace, and VirusTotal scan results \footnote{https://www.virustotal.com/gui/home/upload} (number of antivirus flags). Our focus spanned from 2010 to 2023, specifically targeting marketplaces offering exclusively Android apps, like Google Play Store, Appchina, F-Droid, Lmobile, and Proandroid. Androzoo, with its consistent weekly updates, excels in capturing the evolution over time. Although it doesn't have labels, it compensates by indicating the number of antivirus flags per app, a facet we will discuss in-depth later.
However, data collection was not always smooth sailing. The procedure demanded significant time and occasionally, we stumbled upon apps that were inaccessible in the databases or came with corrupted APK files.

\subsection{Preprocessing and Feature Representation}

The preprocessing phase for all the datasets involved several steps to prepare the APKs for analysis. Initially, the APKs were decompiled using APKTool\footnote{https://apktool.en.lo4d.com}, which systematically arranged the app's files into separate folders. From here, the pivotal Manifest.xml file was extracted from each app. This file is integral as it contains essential details about the app, particularly its permissions. We further refined the data by segmenting the extracted permissions into individual entries.
Consequently, we structured our composite dataset encompassing samples from Drebin, Maldozer, and Androzoo. Each entry in this dataset is characterized by three columns: APK hash name (ID), the processed permissions (Text), and the malware classification (Label). In the case of Androzoo,  the 'Label' column reflects the flag count based on existing VirusTotal reports within the dataset. To further streamline the data, instances from the Androzoo dataset were distributed into distinct CSV files, each consolidating apps from a specific year.

\subsection{Modeling}
\begin{table}
\centering
\caption{\small BERTroid Model Configuration}
  \begin{tabular}{l l }
\hline   Component & Configuration  \\
\hline 
BERT Base Model &	bert-base-uncased \\
BERT Output Used & Pooler Output / Last Hidden State \\
Linear Layer 1 &	Input: 768, Output: 128 with ReLU Activation \\
Linear Layer 2 & Input: 128, Output: 2 \\
Fine-Tuning BERT &	True / False \\
Optimizer &	AdamW with scheduled learning rate \\
\hline
\end{tabular}
  \label{tab1}
\end{table} 
This section introduces the architecture and components of our proposed model, leveraging the BERT language model for binary classification of Android applications into malicious or benign.

\textbf{BERT-based Architecture:} The cornerstone of our proposed model is the BERT language model. Its bidirectional context modeling makes it apt for understanding and capturing intricate contextual relationships in text sequences, such as those found in Android application data where each app is represented by its permissions. The "bert-base-uncased" version of BERT serves as the backbone of our model due to its widespread acceptance and use in the NLP community.
In our context, the Android application data comprises text sequences where every app is characterized by its permissions. This intricate structure makes BERT an apt choice, as it can encode these sequences into dense representations that encapsulate the contextual significance of each permission in relation to others.

\textbf{Neural Network Layers:} On top of the BERT model, two linear layers are added to tailor it for the classification task. The first linear layer is designed to take the 768-dimensional output embeddings from BERT and reduce their dimensionality to 128. This transformation is paired with a ReLU activation function to introduce non-linearity. The second linear layer then further maps these 128-dimensional vectors to a 2-dimensional output. These dimensions correspond to the probability scores for the application being benign or malicious as illustrated in Figure \ref{figappr}.

\textbf{BERT Output Utilization:} An interesting facet of our model lies in the choice between using BERT's "pooler\_output" or its "last\_hidden\_state." The former provides a pooled output, primarily the output of the initial "[CLS]" token. This is particularly useful for sentence-level classification tasks where a holistic representation of the input's semantics is required. Conversely, the "last\_hidden\_state" offers detailed embeddings for each token, encapsulating their individual meanings in the context of the entire input. It's optimal for tasks requiring token-level precision.

\textbf{Fine-tuning and Optimization:} Fine-tuning is an integral part of our modeling strategy. By default, BERT parameters are updated during training, thus tuning it for the specific malware detection task. However, we also offer the flexibility to freeze the BERT parameters and train only the additional layers. For optimization, the "AdamW" optimizer is employed, known for its efficiency and rapid convergence. Alongside, a learning rate scheduler with warmup is utilized to adaptively adjust the learning rate during training, ensuring a smooth training process.

\textbf{Model Configuration Summary:} The various components and configurations of our model are summarized in Table \ref{tab1} for clarity.

\section{Evaluation}\label{sec5}
In this section, we detail BERTroid's experiments and results, accessible in the replication packages\footnote{}. We discuss training hyperparameters, an ablation study on model configurations for malware detection, and address research questions, supplemented by additional analyses for thorough model performance insights.

\subsection{Hyperparameters}
To optimize our BERT-based model for Android malware detection and ensure peak performance, we meticulously considered several essential hyperparameters. We initialized the learning rate to a value of 2e-5, as recommended by BERT's authors. This value serves as a starting point, and the actual learning rate during training is adjusted dynamically using a scheduler to ensure effective convergence. The batch size, set at 16, determines the volume of training data processed concurrently in each iteration. We specify the number of epochs as 5, indicating the total times the entire dataset is processed by the model during training. Additionally, we examined the Pooler Output and the Last Hidden State as alternative methods for obtaining BERT's output. We also explored the implications of fine-tuning, which involves training the entire model, including the BERT layers, on the specific malware detection task. These hyperparameters play a pivotal role in defining the model's performance and efficacy in detecting Android malware.

\subsection{Ablation Study: Effectiveness of Model Configurations}

To understand the individual and collective contributions of different configurations to the effectiveness of our malware detection model, we conducted an ablation study. Specifically, we explored:

\begin{itemize}
\item  BERT's output :  Pooler Output / Last Hidden State.
\item Comparing performance when fine-tuning the entire model, including BERT layers, versus only fine-tuning the succeeding linear layers.
\end{itemize}

For clarity, we designed four experiments, illustrated in Figure \ref{figappr}, to determine the most potent configuration for this task:

\begin{itemize}
\item \textbf{Exp 1}: Pooler output without fine-tuning.
\item \textbf{Exp 2}: Last hidden output without fine-tuning.
\item \textbf{Exp 3}: Pooler output with fine-tuning.
\item \textbf{Exp 4}: Last hidden output with fine-tuning.
\end{itemize}

For the experiments conducted, our primary objective was to discern the performance differences among various configurations. We didn't engage in extensive training with a large dataset or over an extended duration. Instead, the focus was on a concise experiment to gauge the efficiency and effectiveness of each configuration.
To facilitate this, we opted for the Drebin dataset, which is already labeled and comprises both benign and malware apps collected within the same timeframe. From Drebin, the entire malware subset, consisting of 5,560 apps, was utilized. To maintain representation in the benign category, we employed a sample size calculator \footnote{https://www.surveysystem.com/sscalc.htm} and randomly chose 8,911 apps from the vast benign collection. This selection was based on a 99\% confidence level with a 1.3\% confidence interval. Following this, the APK files from the Drebin sample were downloaded and readied for subsequent analysis.
Post preprocessing, our final dataset encompassed 14,366 instances, distributed between 5,538 malware and 8,828 benign apps. This dataset was then employed for meticulous testing of model configurations using K-fold cross validation (K=5). Moreover, a standard split ratio of 80-20\% was maintained for training and testing respectively. The performance results are depicted in Table \ref{tab2}.

\begin{table}
\centering
\caption{\small Performance Comparison of Different Model Configurations }
\resizebox{\linewidth}{!}{
\begin{tabular}{c c c c c c c}

\hline   Experiment  & Test Loss & Accuracy & Precision & Recall & F1-Score & MCC  \\
\hline Exp 1 & 0.635	&0.689&	0.755&	0.769&	0.679&	0.294\\
 Exp 2 &0.233	&0.917	&0.926	&0.917	&0.916	&0.821	\\
 Exp 3&0.101&	0.965&	0.969&	0.965	&0.965&	0.925\\
 Exp 4&0.100&	0.964	& 0.968&	0.963&	0.964&	0.922\\

\hline

\end{tabular}}
 
  \label{tab2}
\end{table}
Table \ref{tab2} showcases the outcomes of four distinct experiments. Exp 1 reports a moderate accuracy of 0.689, complemented by a commendable precision of 0.755 and recall of 0.769. Nonetheless, the F1-Score of 0.679 suggests a slight disparity between precision and recall. The MCC score of 0.294 further indicates that the predictions have a modest correlation with the true labels.
\newline Contrastingly, Exp 2 presents a remarkable enhancement in results, boasting an improved accuracy over Exp 1. Its precision, recall, and F1-Score collectively convey a harmonized model performance. An MCC of 0.821 for Exp 2 signifies a superior alignment of predictions with the true labels than in Exp 1.
\newline Exp 3 stands out further with an accuracy of 0.965, underscored by exemplary precision, recall, and F1-Score values.
\newline Exp 4, resonating with Exp 3, records an accuracy of 0.964, alongside outstanding precision, recall, and F1-Score metrics. The MCC values around 0.92 for both of these experiments attest to the model's robust alignment with the actual labels.
\newline A pivotal observation is the marked escalation in performance transitioning from Exp 1 to Exp 2. This shift, achieved primarily by preferring the last hidden state over the pooler output, accentuates the crucial role of pooling strategy selection for optimizing BERT's capabilities in malware classification. The near-congruent outcomes of Exp 3 and Exp 4 underline the power of the fine-tuning process, which seemingly optimized the model to a stage where further tweaks brought minimal variations. This highlights the pivotal role of fine-tuning in not just amplifying performance, but also ensuring consistent outcomes. The leap in performance between Exp 1 and Exp 2, presumably steered by pooling strategy modifications, stresses the gravity of methodological decisions when harnessing BERT for malware classification tasks. In sum, prioritizing the last hidden output and refining the BERT model markedly amplifies its efficiency in detecting malware within Android platforms.

\fbox{
\begin{minipage}{8cm}
In conclusion, the combination of fine-tuning the BERT model and opting for the last hidden output notably enhances the model's proficiency in malware detection for Android applications. The process of fine-tuning is paramount, fortifying the model's robustness and reliability for the task.
\end{minipage}
}

\subsection{Research Questions}
Following our ablation study, we present the research questions that steered our examination.  In this section, we also detail the evaluation protocols used to rigorously evaluate the model's proficiency and performance.
\begin{itemize}

    \item \textbf{RQ1:} What is the optimal number of antivirus flags to classify an application as malware?
\begin{itemize}
\item \textbf{RQ1.1:} Based on a hybrid manual protocol
\item \textbf{RQ1.2:} Relying on a machine learning model
\end{itemize}
     
     
    \item \textbf{RQ2 :} How does our proposed solution perform compared to state-of-the-art baselines?
    \item \textbf{RQ3 :} How does the resiliency of our approach against permission evolution vary when performing cross-validation over time?
     
\end{itemize}

\subsubsection{Evaluation and results for RQ1}


Effective malware detection in Android security is paramount. Inaccuracies not only expose systems to threats but also erode user confidence. While a higher number of antivirus alerts often suggests an app's potential malicious intent \cite{ref13_zakeya_probing_2022}, the diverse heuristics of antivirus solutions can make these alerts ambiguous.
Adjusting the alert threshold is a double-edged sword. An elevated threshold might overlook malicious apps, whereas a lowered one could falsely label benign apps as threats.
For our research, establishing an optimal threshold serves dual purposes. First, we aim to minimize misclassification risks. Second, we require a reliable threshold to effectively label the Androzoo dataset, enabling us to investigate the temporal evolution of malware behavior in apps. This insight helps us understand how malware strategies have transformed over time, ensuring our model's resilience against evolving threats.
To achieve this, we began with a hybrid manual protocol, leveraging human expertise. The nuances of human judgment in security evaluations offer a robust benchmark. Following this, we turned to our machine learning model to determine a threshold informed by dataset analysis.

\paragraph{\textbf{Evaluation and Results for RQ1.1}}
 
\begin{figure}
    \centering
    \includegraphics[width=80mm]{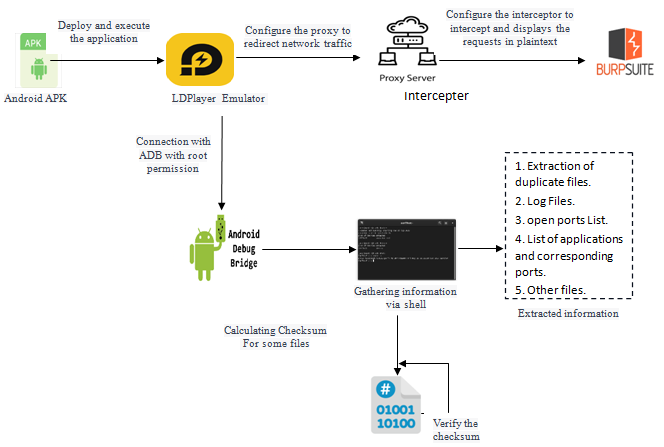}
    \caption{\small Manual analysis protocol for malware detection.}
    \label{fig_aa}
\end{figure}

To achieve this, we employed the Androzoo dataset for our experiments and utilized a manual protocol to investigate app behaviors within a sandbox environment. Figure \ref{fig_aa} provides an illustration of our methodology.
To elucidate our proposed protocol, we detail the most crucial stages in the subsequent steps. Details of the protocol are provided in the replication package:

\renewcommand{\labelenumi}{\arabic{enumi}.}
\begin{enumerate}
\item The Pentester Engineer deploys and runs the application in a Sandbox environment. For this protocol, we have opted to use the LDPPlayer emulator \footnote{https://www.ldplayer.net/} as the sandbox. 
\item After deploying the application on the emulator (Sandbox), we establish a connection with Android Debug Bridge (ABD)\footnote{https://developer.android.com/tools/adb} to the application's host with root privileges. Then, we utilize shell commands to gather information that we consider relevant for detecting malicious behaviors, including : various duplicate files, the application log files, the Android log files, and the list of all open ports along with their associated addresses.


\item Then, for our Android device under test, we configure a proxy (android integrated proxy)  that redirects network traffic to an intermediate interceptor application \footnote{https://portswigger.net/burp} that intercepts and displays the various requests in plaintext. Our goal is to thoroughly review these requests and look for sensitive information,including : credit card details, plaintext passwords (not hashed), data intended for unauthorized parties, or any other information flagged by the experts as a security rule violation.



\item For integrity checking, our protocol uses two kinds of files:
\begin{enumerate}
     \item Firstly, we focus on files that are not meant to be modified by users or regular applications, as altering these files can result in system instability or security vulnerabilities. Some of these critical files include build.prop, hosts and priv-app.
     \item Secondly, we generate test files in various formats commonly utilized by Android malware, including apk, py, docx, sh, and bak files.
\end{enumerate}
Subsequently, we calculate the hash of these files and verify that the hash remains unchanged after running the app. Any alteration in the hash serves as a persistent indicator of malicious behavior.
\end{enumerate}

During our manual review of numerous apps from the Androzoo dataset, we discovered multiple misclassifications by the VirusTotal platform. Several benign applications were mistakenly labeled as malicious. For instance, F-Secure AV Test\footnote{http://surl.li/jqiup} , an app designed to verify antivirus detection capabilities, was flagged as malware by 48 security vendors due to its incorporation of malware signatures for testing. Other apps were wrongly labeled malicious due to intrusive pop-up ads, which, while potentially disruptive, aren't directly harmful.
Delving deeper, one of our most noteworthy findings revolved around an application introduced in 2016. It was tagged as malicious by 16 security providers on the VirusTotal platform. However, our detailed manual inspection revealed that the application was innocuous, displaying no malevolent tendencies. Intriguingly, when we authenticated our findings by re-scanning the application on the VirusTotal platform, the results corroborated our assessment, with no security vendor identifying it as harmful. This stark contrast in classification not only signifies the substantial progress in malicious application detection mechanisms but also highlights that the outcomes vary significantly when the same application is analyzed using updated versions of antivirus software.

We observed that many malware developers utilize a method known as 'APK Embedding' to hide malicious behaviors within a seemingly legitimate application. This method involves incorporating one or more extra APK files, typically carrying malicious intent, into the APK of an apparently benign app using different extensions. After this app gets installed on an Android device, the concealed malicious APK is covertly activated, carrying out its harmful tasks without the user's awareness.

\begin{figure}[htbp]
\centerline{\includegraphics[width=85mm,height=25mm]{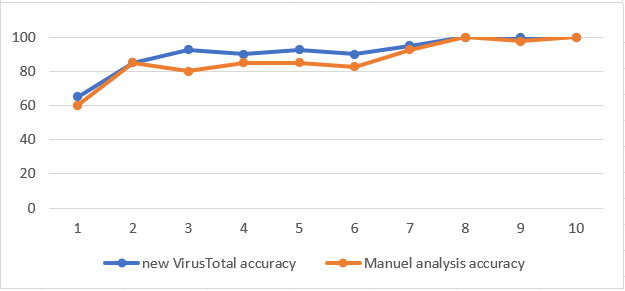}}
\caption{\small Comparing manual and VirusTotal's new classification accuracy against Androzo dataset repoted flags.}
\label{fig_bb}
\end{figure}

Regarding our manual analysis results, we initially examined 106 applications from the Androzo dataset. These apps covered a period from 2016 to 2023 and had varied thresholds ranging from 0 to 48. From our thorough assessment, we found that all apps within the Androzo dataset labeled as malicious with a threshold above 10 aligned with this categorization at a 92.18\% accuracy rate (60 out of 64 applications). On the other hand, apps identified with a security vendor threshold of 10 or below had an accuracy of 85\% (36 out of 42 applications). This data led us to delve deeper into apps detected with 10 or fewer flags. In our subsequent phase, we closely inspected apps within this flag bracket to determine the ideal threshold for malware classification. The findings from our manual analysis are depicted in Figure \ref{fig_bb}. Further, when evaluating apps deemed malicious with thresholds between 11 and 14, we identified consistent outcomes, with minor variations in apps flagged as malicious by eight to ten antivirus solutions.

The graph in Figure \ref{fig_bb} illustrates the detection precision comparison between the content in the Androzoo dataset and the results of our manual analysis compared to the reported results of a new analysis using VirusTotal. The X-axis represents the thresholds set from 1 to 10, while the Y-axis represents the percentage of precision.

For instance,We observe in the figure \ref{fig_bb}  that the results of the manual analyse and the new VirstTotal scan were considerably different from the Androzo categorization for applications classified as malicious by a single antivirus, ranging from 60\% to 65\%.However, for the threshold 5 where we considered all the applications that were classified as malware by 5 antiviruses,we find that only 80\% of them are correctly classified .The percentage of correct classification increases as the number of antivirus agreeing on the application's classification grows, and it becomes nearly constant once applications are approved by eight or more security vendors. This underscores the efficiency of our manual analysis protocol and supports the idea that classifying a software as malicious based on the conclusions of only one or two security vendors is inadequate. In such cases, a more comprehensive evaluation from multiple sources is imperative for a fair and accurate judgment.

After conducting our manual analysis of the applications, we observed that VirusTotal exhibits significant limitations in detecting applications, especially those that conceal their malicious behavior through techniques like 'APK embedding'. It also exclusively focuses on analyzing the current application, neglecting to evaluate the potential for downloading malicious components in the future.

Nonetheless, the manual examination of applications has proven to be highly effective in identifying malicious applications through concrete evidence of their malicious behaviors. Moreover, it can uncover secure applications that may have been incorrectly classified as malicious software, even in the absence of any malicious or suspicious behaviors. Nevertheless, the drawback of this method lies in its time and resource-intensive nature, as well as its reliance on human expertise, making it a costly endeavor.

\fbox{
  \begin{minipage}{8cm} 
After analyzing numerous Android apps across different criteria, the results suggest that if at least eight security vendors flag an app as malicious, it likely is. However, due to VirusTotal's limitations and challenges with manual analysis, we're exploring alternative malware detection methods.
  \end{minipage}
}

\paragraph{\textbf{Evaluation and results for RQ1.2}}
The manual protocol, adeptly merging static and dynamic analyses, provides a detailed methodology for identifying malware. However, it's not without challenges. At the outset, the protocol is notably time-intensive. The thorough examination of each application makes it less feasible for handling vast datasets. Moreover, its hybrid essence demands considerable resources, both in terms of computational power and expertise.

In light of the constraints of the manual protocol, the significance of RQ2 becomes clear. We're inspired by the potential of static analysis, centering on application permissions. These permissions, pivotal for app functionality, can offer deep insights. Deviations or unusual requests in permissions might be telltale signs of malicious intentions.
To make the most of the vast Androzoo dataset, a sound method for categorizing the apps is imperative. In addressing this research question, we delved into a thorough quantitative review of the Androzoo collection, focusing on the distribution of apps based on their antivirus flags from VirusTotal. The data showed a  majority of apps (4,018,515) without any flags, indicating safe applications. The breakdown further highlighted 553,588 apps with a single flag, 207,469 with two flags, 112,088 flagged five times, and 47,541 with ten flags. As the number of flags grew, the count of apps in each segment decreased, culminating in one app that was red-flagged by an astonishing 53 antivirus programs.


Building on the assumption that apps with no antivirus flags are likely safe, and those with several flags may be suspicious or malicious, it's essential to identify a clear threshold for accurate classification in our study. To do this, we systematically explored flag counts. By setting specific thresholds, we trained our model on different data subsets, allowing us to see changes in the results. This method helps pinpoint the best threshold for separating benign apps from malicious ones.

For these tests, we used a sample size calculator set at a 99\% confidence level and a 1.5\% margin of error, ensuring our results are statistically sound. Table \ref{tab3} displays the various thresholds used in each test, as well as the total instances from the Androzoo dataset and the chosen sample size. For example, in the experiment 1, apps with 0 flags were labeled as safe, while those with one or more flags were considered malicious. By the experiment 9, apps with 0 flags remained categorized as safe, but only those with ten or more flags were deemed malicious.

Overall, we carried out 9 experiments, each with different flag threshold criteria. Table \ref{tab3} lists the total number of begnin and malicious apps for each experiment from the Androzoo dataset, matched with the sample size recommended by the calculator. Our goal is to discover the most effective threshold for classifying Android apps. We then trained on labeled data according to each experiment's criteria and tested on validated app samples from the previously mentioned Drebin and Mladozer datasets.

\begin{table}
\centering
\caption{\small Threshold Experiments and Sample Sizes for Android App Classification}
\resizebox{\linewidth}{!}{
  \begin{tabular}{c c c c c }
\hline  Experiments& Begnin & Malware & Begnin & Malware  \\
& flag - size &flag - size &Sample size&Sample size\\
\hline   Exp 1  & 0 (4018515) &  1+( 1989169) & 7382 &7369\\
Exp 2  & 0-2 (4779599) &  3+ (1228085) & 7385 &7352\\
Exp 3  & 0-5 (5128766) &  6+ (878918) & 7385 &7334\\
Exp 4  & 0-7 (5327425) &  8+ (680259) & 7386 &7316\\
Exp 5  & 0-9 (5404547) &  10+ (603137) & 7386 &7306\\
Exp 6 & 0-11 (5555272) &  12+ (452412) & 7386 &7277\\
Exp 7  & \textbf{0} (4018515) &  \textbf{6+} (878918) & 7382 &7334\\
Exp 8  & \textbf{0} (4018515) &  \textbf{8+} (680259) & 7382 &7316\\
Exp 9 & \textbf{0} (4018515) & \textbf{ 10+} (452412) & 7382 &7277\\  

\hline

\end{tabular}}
 
  \label{tab3}
\end{table}

The results of the experiments presented in Table \ref{tab33} offer valuable insights into the selection of an appropriate threshold for laveling Android applications as malware. In contrast to the findings of the study \cite{ref13_zakeya_probing_2022}, experiment 1, with a threshold of 1+, showed the lowest accuracy and MCC values, indicating that not all applications flagged by at least one antivirus are genuinely malicious. 
\newline Moving from experiment 1 to experiment 9, there is a consistent trend of performance improvement. As the threshold values increase, accuracy, precision, recall, and F1-score all exhibit an upward trend. This suggests that adopting a more conservative approach by raising the threshold for classifying apps as malware results in better overall performance, particularly in correctly identifying malware samples.
\newline However, we observed that performance declined when the threshold was set beyond 12, presented as experiment 6. To further refine the threshold selection, we conducted experiments with combined thresholds (experiments 7, 8, and 9). In these experiments, we set the benign threshold at 0 flags and varied the malware threshold between 6+, 8+, and 10+ flags, as they demonstrated promising performance with unique thresholds.
\newline Notably, experiment 8 stood out with the highest performance. By classifying apps with 0 flags as benign and apps with 8 or more flags as malware, it achieved the highest values for accuracy, precision, recall, F1-score, and MCC, with respective values of 0.929, 0.936, 0.929, 0.929, and 0.858. This particular threshold configuration strikes an excellent balance between correctly identifying malware samples (high recall) and minimizing false positives (high precision).
The Matthews correlation coefficient (MCC) corroborated this trend, as experiment 8 obtained the highest MCC value, further highlighting the effectiveness of this threshold configuration in improving overall malware detection performance.

\begin{table}
\centering
\caption{\small Performance Evaluation of Different Thresholds for Android Malware Classificationn}
\resizebox{\linewidth}{!}{
  \begin{tabular}{c c c c c c  }
\hline Experiment	& Accuracy&	Precision	&recall	&F1-score&	MCC	\\
\hline Exp 1	&0.892	&0.906	&0.892	&0.892	&0.786	\\
Exp 2	&0.913	&0.923	&0.913	&0.913	&0.826	\\
Exp 3	&0.902	&0.914	&0.902	&0.902	&0.805	\\
Exp 4	&0.915	&0.926	&0.915	&0.915	&0.829	\\
Exp 5	&0.910	&0.921	&0.910	&0.911	&0.822	\\
Exp 6	&0.907	&0.919	&0.907	&0.908	&0.816	\\
Exp 7	&0.925	&0.933	&0.925	&0.926	&0.850	\\
Exp 8	&0.929	&0.936	&0.929	&0.929	&0.858	\\
Exp 9	&0.925	&0.934	&0.925	&0.925	&0.852	\\
\hline

\end{tabular}}
 
  \label{tab33}
\end{table}
The alignment of our findings, where both the machine learning model and the manual protocol converge at the threshold of eight, is emblematic of our model's robustness. This parallel underscores the efficacy of machine learning, suggesting that even when based purely on static analysis, valuable insights can be derived, particularly with permissions acting as indicative markers of app behavior. Although dynamic analysis dives deeper, it's not without its vulnerabilities. Tactics like time bombs and various evasion methods are employed by developers to circumvent detection. These findings champion the machine learning approach to static analysis, emphasizing that while it may not delve as deep as dynamic analysis, it stands strong as a time-efficient and robust methodology for malware detection.

\fbox{
  \begin{minipage}{8cm} 
    The results emphasize the importance of selecting a suitable threshold to achieve precise classification. Among the experiments conducted, Experiment 8, with a combined threshold of 0 flags for benign and 8+ flags for malware, emerged as the optimal choice, delivering a well-balanced performance across various evaluation metrics. 
  \end{minipage}
}

\subsubsection{Evaluation and results for RQ2}
In this section, we address our research question concerning the effectiveness of our approach in detecting Android malware applications in comparison to state-of-the-art baselines. We conducted a comprehensive evaluation of our model, exploring various datasets combinations and hyperparameter settings to identify the most optimal configurations. To illustrate the comparison, we present Table \ref{tab5}, which offers a detailed analysis of our model's performance alongside that of the state-of-the-art baselines.

\begin{table}
\centering
\caption{\small Performance Comparison with State-of-the-Art Baselines }
\resizebox{\linewidth}{!}{
  \begin{tabular}{ l c c c c c }
\hline  Approach &	Accuracy & 	Precision &	Recall & F1-score & MCC\\
\hline  Androvul\cite{ref13_zakeya_probing_2022} &  0.92 & -  & - & 0.87 & - \\
        Maldozer\cite{ref8_karbab_maldozer_2018} & - & 0.9984& 0.9984 & 0.9984 & - \\
        MalBERT\cite{ref9_9659287} & 0.9761 & -& - & 0.9547 & 0.9559 \\
        DistilBERT\cite{ref9_9659287} & 0.9542 & -& - &  0.9542 &  0.9087 \\
        RoBERTA \cite{ref9_9659287}& 0.9533 & -& - &  0.9499 &  0.907 \\
        \textbf{BERTroid} & \textbf{0.9974}	& \textbf{0.9988}&	\textbf{0.9987}&	\textbf{0.9987}&	\textbf{0.9987}\\

\hline

\end{tabular}}
 
  \label{tab5}
\end{table}
The comparative analysis of the Table \ref{tab5} results indicates that our proposed approach, BERTroid, outperforms all the state-of-the-art baselines in detecting Android malware applications. BERTroid achieves a significantly higher Accuracy of 0.997416 and F1-score of 0.9987, surpassing the other baselines. Moreover, BERTroid demonstrates exceptional Precision (0.9988) and Recall (0.9987) values, indicating a balanced performance in identifying true positives and minimizing false negatives. Additionally, the high MCC value of 0.9987 highlights the strong correlation between BERTroid's predictions and the true labels. These findings emphasize the superiority of BERTroid and its potential to be an effective and reliable solution for Android malware detection.

To further evaluate the effectiveness of BERTroid, particularly in the context of Androzoo experiments, we present Table \ref{tab6} that provides a comparative analysis of the results obtained by the Androvul approach \cite{ref13_zakeya_probing_2022} and our BERTroid model when applied to the same Androzoo dataset, using the same threshold experimentations.

\begin{table}
\centering
\caption{\small Performance Comparison with Androvul detailed experimentation }
\resizebox{\linewidth}{!}{
  \begin{tabular}{ l c c c c c c c}
\hline Experiment  & Approach & F1-score & AUC-ROC\\
\hline  Exp 1  &  Androvul   & 0.80 &0.85\\
0 Begnin - 1+ Malware & BERTroid &	0.8926	&	0.8941\\

\hline  Exp 2  &  Androvul & 0.82& 0.87\\
0-2 Begnin - 3+ Malware & BERTroid &	0.9136&	0.9136\\
\hline  Exp 3  &  Androvul   & 0.89  & 0.83\\
0-9 Begnin - 10+ Malware & BERTroid &	0.9111&	0.9141
\\

\hline

\end{tabular}}
 
  \label{tab6}
\end{table}
According to Table \ref{tab6}, in Experiment 1, we exhibit higher F1-score and AUC-ROC values compared to Androvul. BERToid also achieves an accuracy of 0.89. A similar trend is observed in Experiment 2, where BERTroid demonstrates superior performance in all evaluation metrics compared to Androvul. In Experiment 3, BERTroid maintains its dominance with higher values of evaluation metrics . Overall, the results consistently show that BERTroid outperforms Androvul across all experiments, highlighting its effectiveness in detecting Android malware applications and justifying its superiority as a robust and reliable approach.

\fbox{
  \begin{minipage}{8cm} 
    In comparison with state-of-the-art baselines and a detailed evaluation against the Androvul approach, BERTroid model consistently demonstrates superior performance in detecting Android malware applications.
  \end{minipage}
}

\subsubsection{Evaluation and results for RQ3} 
\begin{figure}
    \centering {\includegraphics[width=7cm, height=90mm]{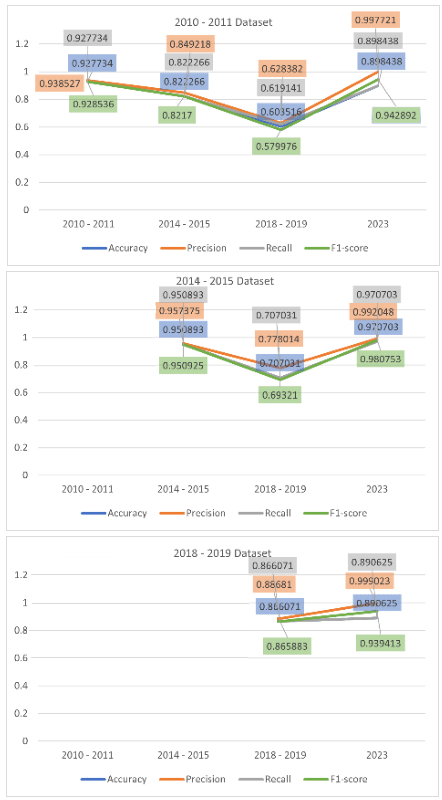}}
    \caption{\small Cross Time validation results}
    \label{fig5}
\end{figure}
In this section, we seek to assess our model's resilience against the evolving landscape of Android app permissions over time. As app behaviors and requirements evolve, their permission combinations change, potentially affecting malware detection model accuracy. Upon analyzing the Androzoo dataset, we observed that yearly shifts in permission combinations were not significantly impactful. However, with a gap of three years, many new permissions and combinations emerge. Hence, we focus on the dataset from years 2010, 2011, 2014, 2015, 2018, 2019, and 2023, testing at 3-year intervals to encapsulate these evolutionary shifts. Our experiments involve training the model on 2010-2011 data, then testing against subsequent intervals. This process is repeated with 2014-2015 and 2018-2019 datasets. By training on historical data and testing on more recent samples, we aim to comprehend how these evolving permissions influence our model's predictive capabilities over time.

Figure \ref{fig5} highlights discernible patterns. When trained on the 2010-2011 dataset and tested on 2014-2015 and 2018-2019, there was a significant decrease in accuracy. Similarly, training on the 2014-2015 dataset and testing on 2018-2019 exhibited recurring patterns across different training and testing phases, predominantly for the years 2018-2019 and 2023.

For 2023, the results stood out, potentially due to a reduced number of antivirus flags, evidenced by just 11 flags for distinct apps. This reduction suggests a possible decrease in the diversity of permission combinations for apps that year. However, it's crucial to mention that the data for 2023 is ongoing, and might not fully encapsulate the applications of that year.

For the year 2018-2019, there was a consistent drop in performance across all training scenarios. This year observed a heightened number of attacks, increasing by up to 350\% compared to earlier periods \cite{rr3}. This downturn might be due to the more intricate and diverse app permission combinations introduced during that year. Additionally, 2018 saw the rollout of Android Pie (Android 9)\footnote{https://developer.android.com/about/versions/pie/android-9.0-changes-all}. This version introduced numerous security and privacy enhancements, such as robust defenses against malicious apps, the Android Protected Confirmation for sensitive transactions, and encrypted backups secured by a client-side secret \cite{rr2}. Furthermore, Android Pie redefined permission parameters, restricting access to call logs and SMS unless designated as the default handler and limiting the use of some non-SDK interfaces \cite{rr1}.
In the realm of Android, changes in permissions have substantially influenced the ways apps interact. This evolution stems from factors like updates in API levels, leading to permissions being added or removed. Notably, Android 6.0 introduced a dynamic permission model, facilitating on-the-fly permission requests \cite{pp3}. As a result, diverse permission patterns emerged, influenced by user interactions and specific app use cases. This evolution suggests notable permission differences between 2010-2014 and later years. As the Android platform developed, apps started to request permissions in a more sophisticated manner. A model based on older data may face challenges with newer permissions, a sentiment echoed in various security studies\cite{pp1}\cite{pp2}.
\newline Nevertheless, our model showcases remarkable stability, particularly when juxtaposed with the study by \cite{ref8_karbab_maldozer_2018}. Their research noted a 29.34\% performance degradation over two years (2014 training data versus 2016 testing data). In contrast, our model, spanning a four-year interval, recorded merely a 26.31\% decline (2014 training versus 2018 testing). These findings highlight our appraoche's superior resilience and capability compared to other models.

In our research's concluding phase, we emphasized the importance of manual validation to bolster our model's accuracy and reliability. We randomly chose a subset of false positives and negatives to determine their actual status and understand the reasons for their misclassifications through manual review. Upon examination, about 17\% of the apps that our model identified as malware, yet were labeled benign in the dataset due to flag counts, indeed showed malicious tendencies. Still, our model erroneously marked some apps as malicious. The following are reasons for these discrepancies:

 \begin{itemize}
     \item Zero-Day Threat: The malicious app may employ new and sophisticated evasion techniques that have not yet been added to the antivirus databases.
     \item Encrypted Payloads: Malicious apps utilize encrypted payloads and obfuscation techniques for concealment.
     \item False Negatives: Antivirus programs aim to minimize false positives (flagging benign files as malicious) and might occasionally produce false negatives (failing to detect malicious files). 
\end{itemize}

With regard to false negatives,our manual validation confirmed that some applications classified as benign by our model were actually malicious.We explain the poorness of the classification by :

 \begin{itemize}
     \item According to our approach, the limited number of permissions required by certain applications keeps them free from suspicion.
     \item Begnin applications embed camouflaged malicious APK files to bypass any static analysis.
     \item Applications that require a new download and installation once they are executed.
\end{itemize}

Drawing from a pivotal study on Android security permissions, it's evident that while permissions have evolved with every Android OS version, they remain a crucial tool for malware detection. This evolution, introducing new permissions and deprecating others, influences the permissions sought by both malicious and benign apps, introducing the challenge of concept drift in machine learning models. Using this foundational knowledge, our model, centered around permissions, demonstrates significant resilience against these temporal shifts, emphasizing the robustness of permissions-focused models in navigating the ever-changing Android landscape\cite{ppp}.

\fbox{
\begin{minipage}{8cm}
Our model demonstrates notable robustness across prolonged intervals relative to alternative systems. Nevertheless, the constrained annual dataset underscores the imperative of sustained data accumulation to further amplify the efficacy of our approach over the years.
\end{minipage}
}

\section{Threats to validity}\label{sec6}
Regarding internal validity, we need to consider the possibility of coding errors or bugs in our model's implementation. To address this concern, we took extensive measures, such as conducting code reviews and rigorous testing throughout the development process. Additionally, we ran multiple iterations of experiments to ensure the consistency and reliability of our results.
To enhance external validity, we carefully selected a diverse range of datasets from various sources, including Drebin, Maldozer, and Androzoo, to encompass a wide array of Android applications in order to evaluate our model's performance across different scenarios. Despite these efforts, it's important to acknowledge that the generalizability of our findings may be influenced by factors such as variations in the Android ecosystem, the prevalence of malware, or the specific datasets used.
Moreover, when comparing our approach with state-of-the-art solutions, we made an effort to select their best reported values for each evaluation metric to ensure a fair and meaningful comparison. However, it's essential to consider that differences in data selection and preprocessing may introduce some uncertainty when interpreting the results.

\section{conclusion}\label{sec7}
In conclusion, our paper presents a comprehensive study on Android malware detection, showcasing our approach's superiority over state-of-the-art solutions in accuracy, precision, recall, F1-score, and MCC. We address the challenge of labeling the vast Androzoo dataset, introducing a manual protocol that significantly enhances accuracy and reliability. Thorough testing on multiple datasets, including Androzoo, Drebin, and Maldozer, validates the model's consistent performance across various scenarios. We also examine our model's resiliency against permission evolution over time, highlighting the need for continuous data collection and model refinement. Ultimately, our proposed manual protocol empowers deeper analysis of the Androzoo dataset, aiding in identifying false classifications and gaining crucial insights into the model's performance.

To enhance the model's ability to identify advanced and newly developed malware threats, future research will focus on incorporating diverse feature sets and data sources. Testing the model across different Android devices and versions will provide insights into its real-world applicability. We are also considering the exploration of ensemble learning strategies by combining predictions from multiple models. Furthermore, we aim to classify malware based on its specific families. Advancements in these domains will further bolster the reliability and efficacy of Android malware detection.
\bibliographystyle{ieeetr}
\bibliography{references}

\end{document}